\documentclass{icrc2009}

\usepackage{graphicx}   % for including figures
\usepackage[caption=false]{caption}    % for captions
\usepackage[font=footnotesize]{subfig} % subfig.sty for a double column floating figure using two subfigures
\usepackage{fixltx2e}
\usepackage{url}

\newcommand{\shorttitle}[1]%
{\markboth{Proceedings of the 31\MakeLowercase{$^{st}$} ICRC, {\L}\'{o}d\'{z} 2009}{#1} }
\newcommand{\etal}{\MakeLowercase{\textit{et al. }}} % "et al."

\begin{document}
\title{The Advanced Gamma-ray Imaging System (AGIS): Simulation Studies}

\author{\IEEEauthorblockN{Gernot Maier\IEEEauthorrefmark{1} for the AGIS Collaboration\IEEEauthorrefmark{2} 
			 }
                            \\
\IEEEauthorblockA{\IEEEauthorrefmark{1}Department of Physics, McGill University, H3A 2T8 Montreal, QC, Canada (maierg@physics.mcgill.ca)}
\IEEEauthorblockA{\IEEEauthorrefmark{2}http://www.agis-observatory.org}
}

% please write the preseter's name and short title (3-4 words maximum)
%    which will appear at the header of the even pages.
\shorttitle{G.Maier \etal AGIS: Simulation studies}
\maketitle

\begin{abstract}

The Advanced Gamma-ray Imaging System (AGIS) is a next-generation ground-based gamma-ray observatory 
being planned in the U.S. The anticipated sensitivity of AGIS is about one order of magnitude better than the
sensitivity of current observatories, allowing it to measure gamma-ray emmission from a large number of Galactic
and extra-galactic sources.  We present here results of simulation studies of various possible designs for AGIS.
The primary characteristics of the array performance - collecting area, angular resolution, background rejection,
and sensitivity - are discussed. 
 \end{abstract}

\begin{IEEEkeywords}
gamma-ray observatory; new instruments; simulations
\end{IEEEkeywords}
 
\section{Introduction}

Ground-based gamma-ray astronomy has seen a revolution in the present decade with the 
operation of new instruments like H.E.S.S., MAGIC, and VERITAS. 
The impressive discoveries of these instruments have demonstrated the scientific potential
of the field.
This rapid progress is expected to continue with the recently launched Fermi Space Telescope 
and the 
next generation of gamma-ray observatories.
The new ground-based detectors will consist of large arrays of imaging atmospheric telescopes (IACTs),
forming instruments with about 10 times the sensitivity of current-generation gamma-ray observatories,
an energy range from 10 GeV to 100 TeV, and a much improved angular resolution of 
1-3 arcmin.
 Two such observatories are currently in development:
 the Cherenkov Telescope Array (CTA\footnote{http://www.cta-observatory.org}) in Europe and the US-led 
 Advanced Gamma-ray Imaging System (AGIS\footnote{http://www.agis-observatory.org}).

\begin{figure}[!t]
\centering
\includegraphics[width=2.5in]{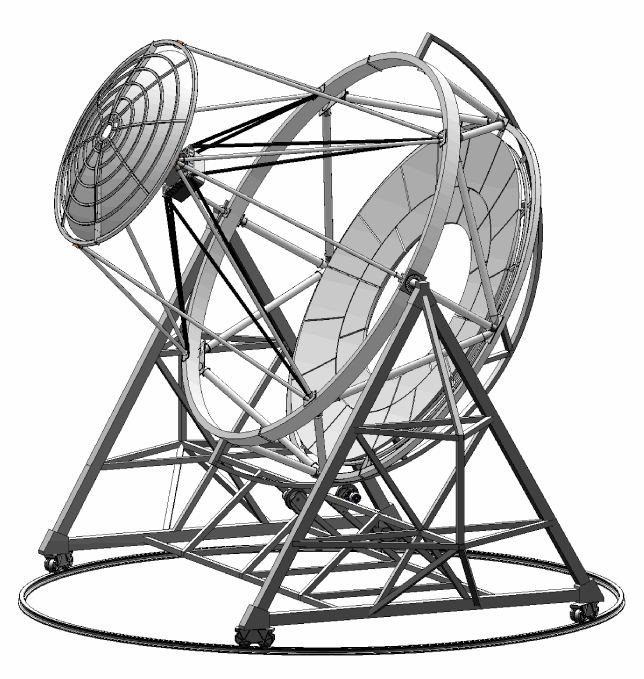}
\caption{Model of a AGIS Schwarzschild-Couder telescope and its two-mirror aplanatic optical system.}
\label{fig:SD}
\end{figure}

AGIS will consist of a large array of medium sized imaging atmospheric
Cherenkov Telescope covering an area of about 1 km$^2$ on the ground.
The conceptional design for AGIS is an array of 36 wide field-of-view
(FOV; $8^{\mathrm{o}}$) telescopes, each of which has an effective
light collection area of 100 m$^2$.  The wide FOV of the AGIS
instruments is motivated scientifically, e.g.,~to provide a
sensitive
Galactic plane survey and the mapping of extended sources and is an
integral part of the array design.  The wide FOV improves the
background rejection
and angular resolution, because it increases the
number of telescopes participating in the
event reconstruction.  This
improves the amount of information available for recognizing
the
nature of the primary (gamma-ray or cosmic ray) and also significantly
improves the angular
resolution (illustrated in Figure
\ref{fig:groundView}).  The wide FOV, large array design of
AGIS
provides significant improvements over a simple
$\sqrt N$
extrapolation ($N$= telescope number)
that one would expect from
simple scaling from current small arrays (VERITAS, H.E.S.S.).  Table
\ref{tab:specification} lists the preliminary specifications of the 36
element AGIS array.

 \begin{table}[!h]
  \caption{Specifications of the 36 element AGIS array.}
  \label{tab:specification}
  \centering
  \begin{tabular}{|l|l|}
  \hline
    & Target \\
   \hline 
  Telescope Spacing & 100 - 150 m \\
  Effective Mirror Area per Telescope & 100 m$^2$ \\
  Field of View (FOV) & 8 deg \\
  Pixelation & 0.05 - 0.10 deg \\
  Effective Collection Area & 1 km$^2$ \\
  Energy Threshold\footnote{The energy threshold is defined as the peak of the differential counting rate
for a Crab Nebula-like spectrum.} & 100 GeV \\
Angular Resolution & 0.02 - 0.05 deg \\
  \hline
  \end{tabular}
  \end{table}

AGIS uses a two-mirror telescope with a Schwarzschild-Couder optical system
\cite{VVV-2007} \cite{Fegan-2007} combined with a compact modular camera (see Figure \ref{fig:SD}).
This design achieves an excellent point-spread function across the wide FOV
and allows at the same time a considerably shorter focal length than traditional 
IACTs.
The compact camera will be equipped with small-sized, integrated photo-sensors 
such as multianode photomultiplier tubes.

\section{Simulations}

Detailed
simulation studies for optimizing
the key performance parameters of the
AGIS array are currently in progress;
first results are given here. 
The computations
include full air-shower simulations
based on CORSIKA \cite{Heck-1998} \cite{Bernlohr-2008}
and a complete model of the
optical and electronic response of the
telescopes\footnote{http://www.physics.utah.edu/gammaray/GrISU/}.
The simulations are built on established software tools that have
been successfully used for characterizing
the VERITAS performance (see e.g. \cite{Maier-2007}).

The key features of the wide-field-of-view AGIS telescopes, i.e. excellent optical 
properties and fine pixelation, are taken into
account. 
The simulations show, for example, that an optical point spread function of 0.05$^{\mathrm{o}}$ or
better can be achieved at any point in the focal plane.

The analysis steps incorporated in the simulation include pedestal calculation, image 
 cleaning, image parameterization, source reconstruction, and calculation of mean-scaled variables
for (cosmic-ray) background suppression, similar to those used in the analysis of VERITAS
data (see e.g. \cite{Acciari-2008}).
The analysis of AGIS events is not yet optimized for the larger
number of pixels and telescopes, and therefore, our results are a conservative estimate and
are likely to improve.

\begin{figure}[!t]
\centering
\includegraphics[width=3.0in]{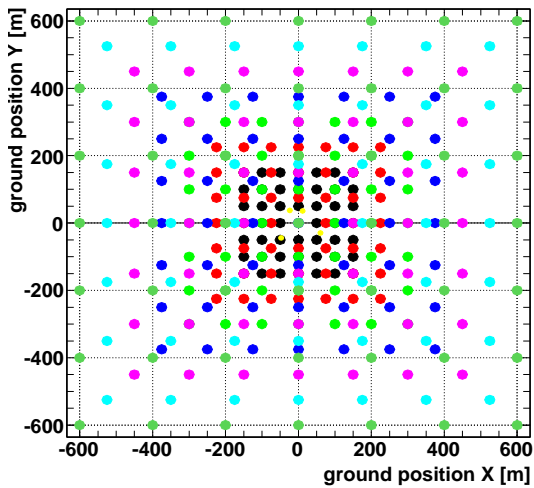}
\caption{Telescope positions used in the CORSIKA simulations.
}
\label{fig:hyperArray}
\end{figure}

Different AGIS configurations with pixel sizes between 0.05 deg and 0.20 deg and telescope
distances between 50 m and 200 m have been simulated.
In order to use computing power most effectively, a large library of simulated CORSIKA events
 with a 'hyper-array' configuration is currently being generated (see Figure \ref{fig:hyperArray}).
 Sub-arrays are selected from this hyper-array in order to optimize angular
resolution, effective area and background suppression. 

\begin{figure*}[!t]
   \centerline{\subfloat[Case I]{\includegraphics[width=3.0in]{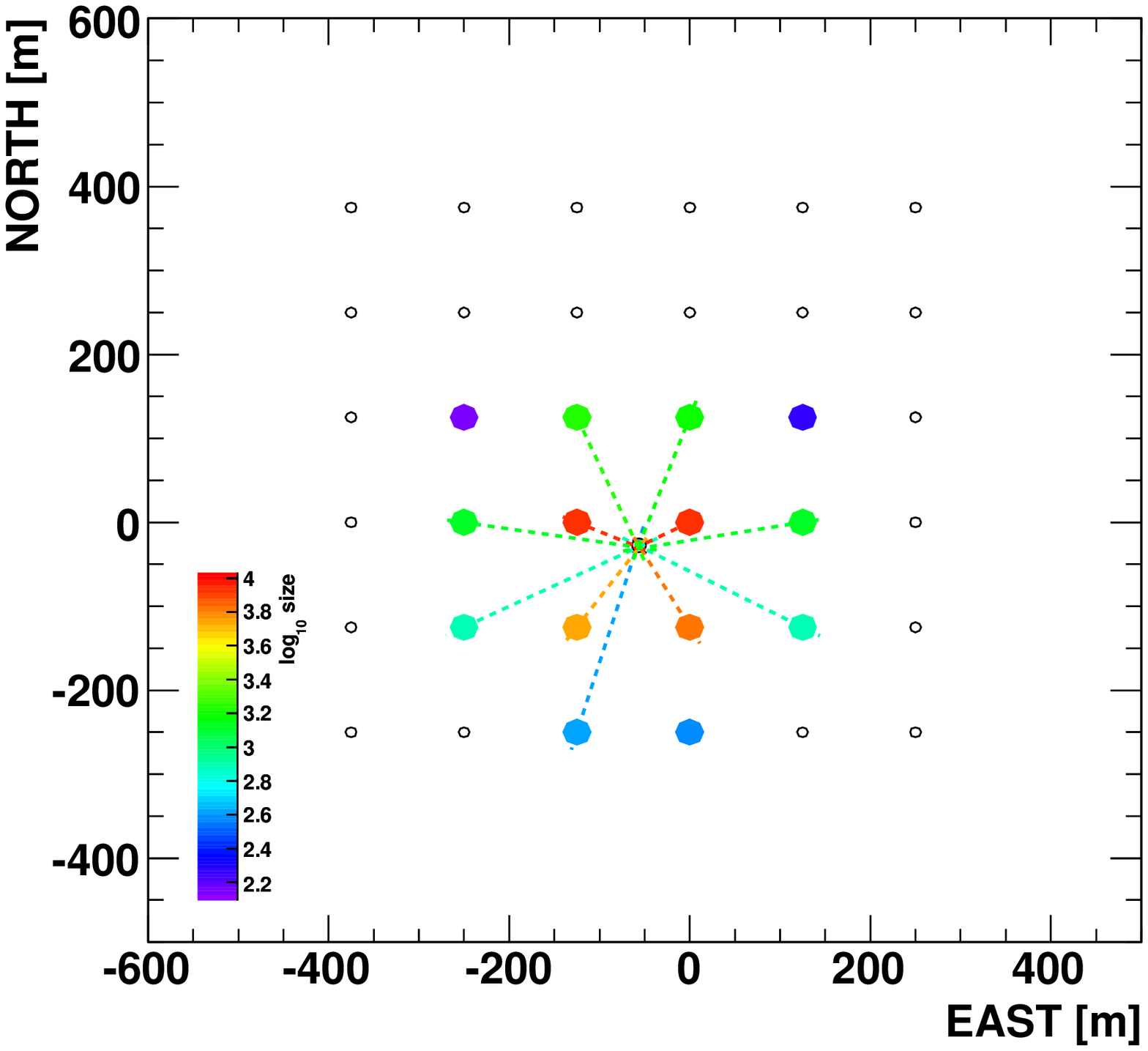} \label{sub_fig1}}
              \hfil
              \subfloat[Case II]{\includegraphics[width=3.0in]{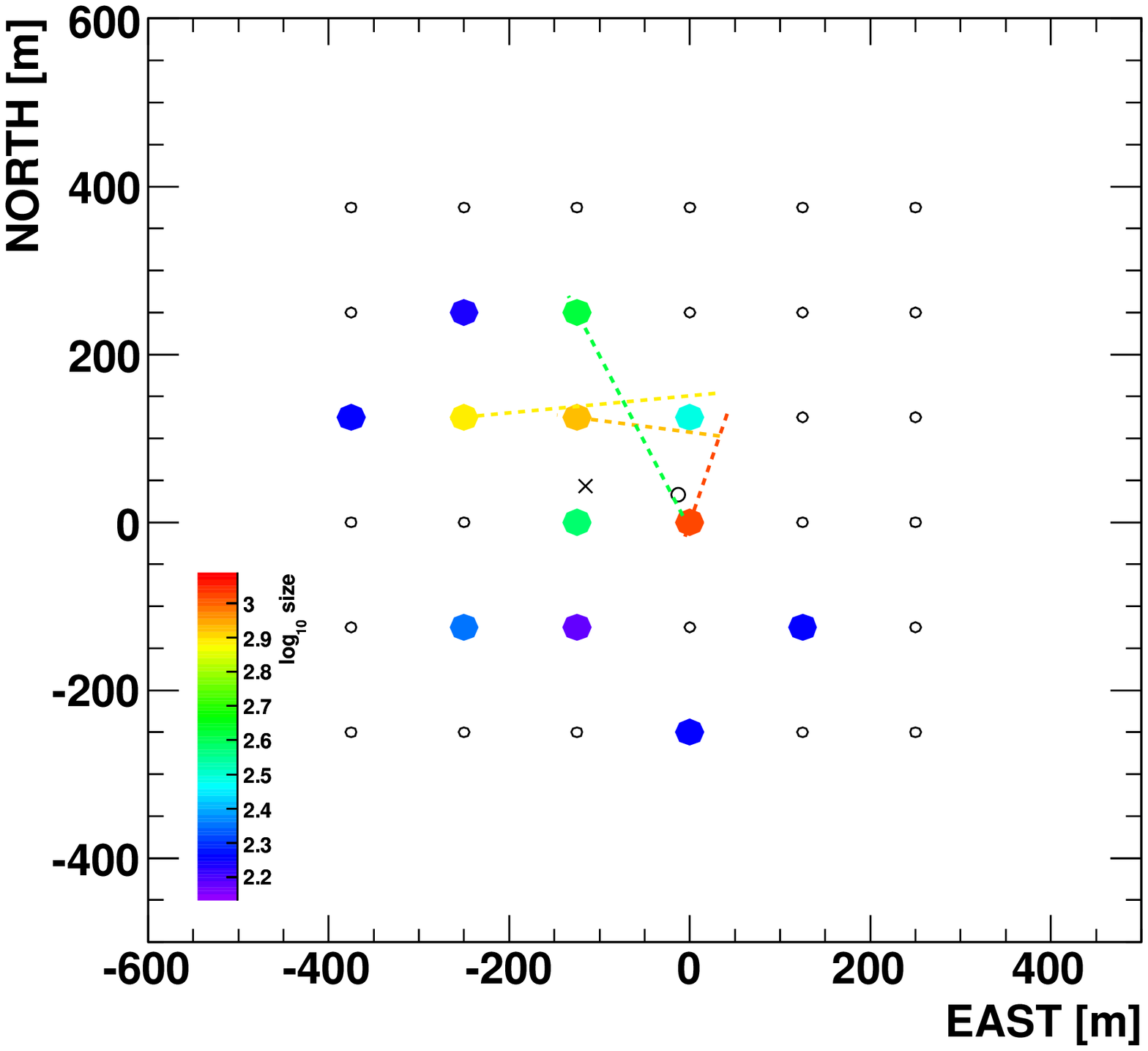} \label{sub_fig2}}
                 }
                  \centerline{\subfloat[Case I]{\includegraphics[width=3.0in]{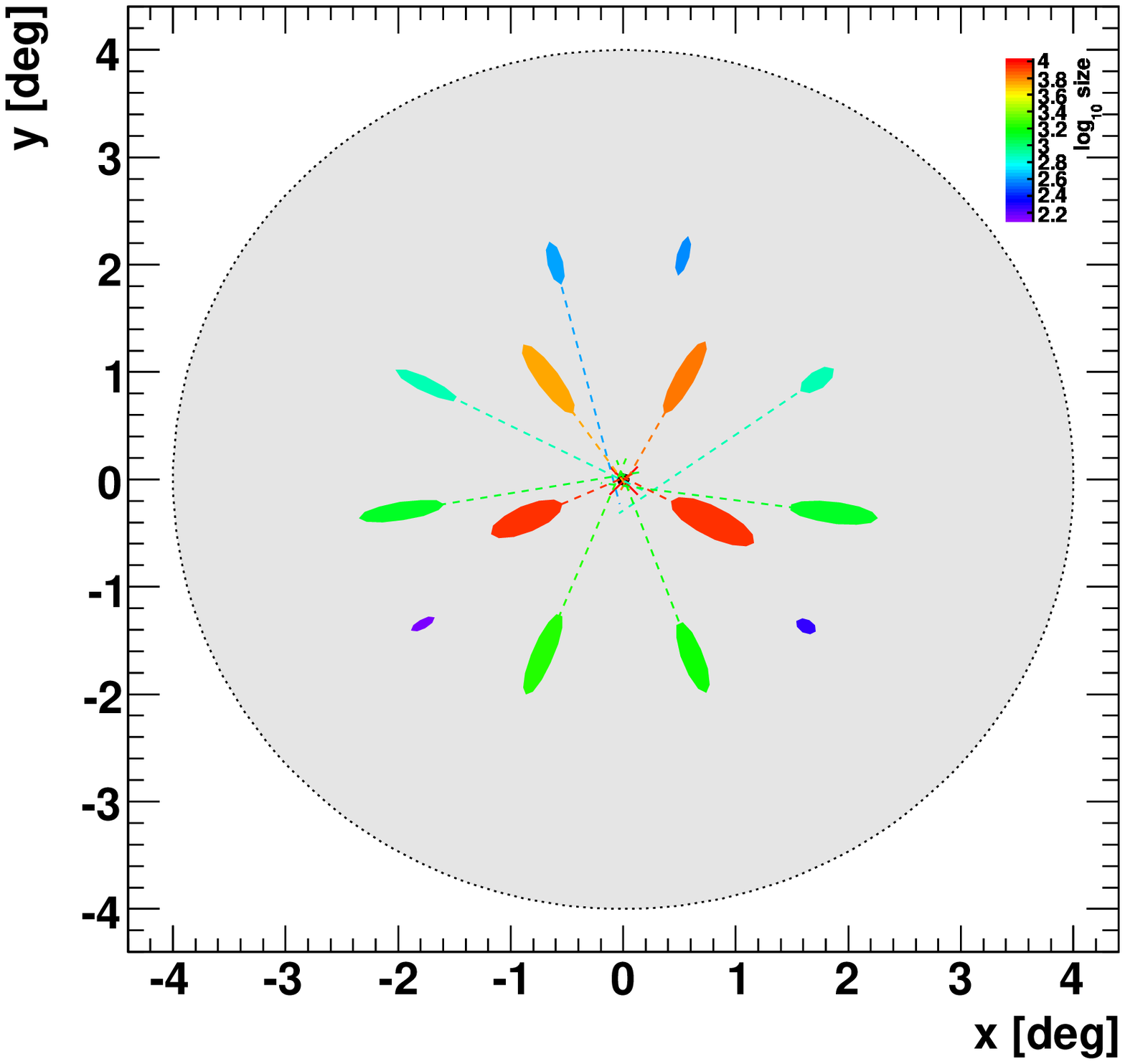} \label{sub_fig3}}
              \hfil
              \subfloat[Case II]{\includegraphics[width=3.0in]{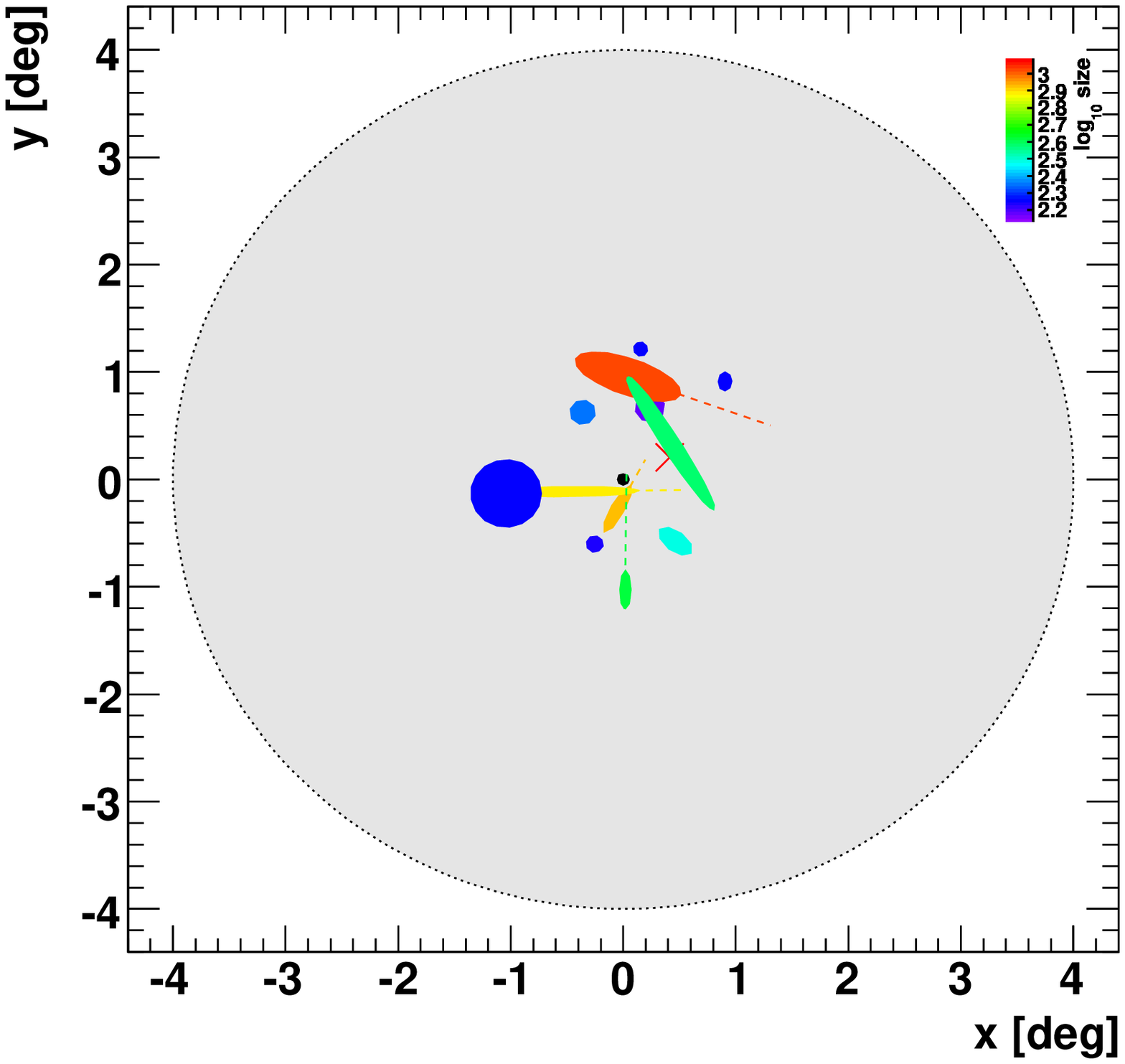} \label{sub_fig4}}
             }
   \caption{
   Detection of a 1 TeV gamma-ray (left figures) and 1 TeV proton shower (right figures) with an array consisting
     of 36 wide-FOV telescopes. 
     Top: Illustration of the shower core reconstruction
     (125 m distance between the telescope, which are drawn as 
     black circle. The position of the shower core is indicated by black crosses.)
     Bottom: Superimposed camera images of the shower from all triggered telescopes.
     The color code indicates the total number of Cherenkov photons measured by each telescope,
     the dashed lines show the orientation of the long axes of the images used for core and direction reconstruction.
     }
   \label{fig:groundView}
  \end{figure*}

Figure  \ref{fig:groundView} illustrates the difference between
a typical gamma-ray and a cosmic-ray induced air shower.
 While a gamma-ray event exhibits a
consistent image pattern with many telescopes contributing to the reconstruction of the arrival
direction, hadronic showers have large fluctuations in their images in different cameras,
providing additional background suppression capabilities. 
 Quantitative studies are underway
but require very large numbers of hadronic simulations because few events actually pass the
primary gamma-ray selection cuts. 
 This already indicates the power of the stereoscopic imaging
technique with large arrays.

\begin{figure}[!t]
\centering
\includegraphics[width=3.0in]{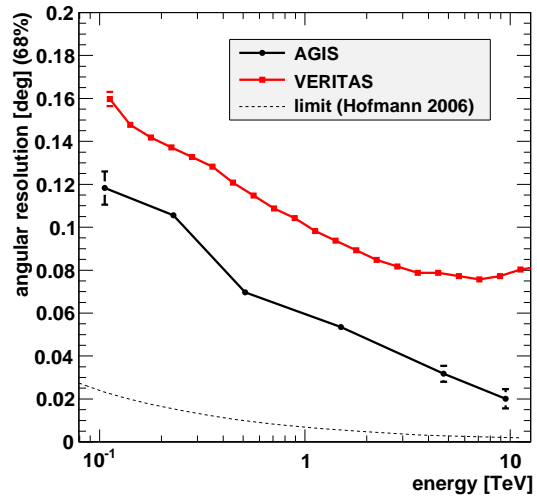}
\caption{Angular resolution (with 0.1 deg camera pixels) vs energy for a 36 telescope array with 
telescope distances of 125 m compared with VERITAS \cite{Holder-2008} and the theoretical limit 
as derived in \cite{Hofmann-2005}.}
\label{fig:AngRes}
\end{figure}

\begin{figure}[!t]
\centering
\includegraphics[width=3.0in]{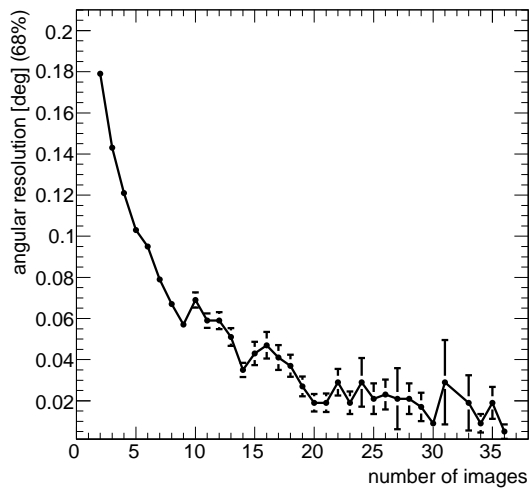}
\caption{Angular resolution vs number of images available in the reconstruction at 1 TeV
for a 36 telescope array with a telescope distance of 125 m (with 0.1 deg camera pixels).} 
\label{fig:AngResN}
\end{figure}

In addition, by measuring the same shower with many telescopes, a 2-3
 times better
angular resolution can be achieved for gamma-ray events
 compared to current small arrays of
IACTs.  Figure \ref{fig:AngRes}
 shows the angular resolution for a 36-telescope array with VERITAS-type
telescopes.  The improvement in angular resolution compared to
 VERITAS is due to the
larger number of telescopes only, see Figure
 \ref{fig:AngResN} and \cite{Hinton-2009}.  It should be noted that
 the calculation of the theoretical limit for the angular
 reconstruction indicated in Figure \ref{fig:AngRes} suggests that an
 order of magnitude better
angular resolution could be possible at
 higher energies.  However, this estimate requires
$\approx10$\%
 coverage at the light pool which is impractical while simultaneously
 trying to optimize
the array size.  While the full improvement may
 not be attainable with a real-world instrument,
it may be possible to
 get close to that at multi-TeV energies or at lower energies with a
 ``graded'' array composed of telescopes of several different sizes.
 
\begin{figure}[!t]
\centering
\includegraphics[width=3.0in]{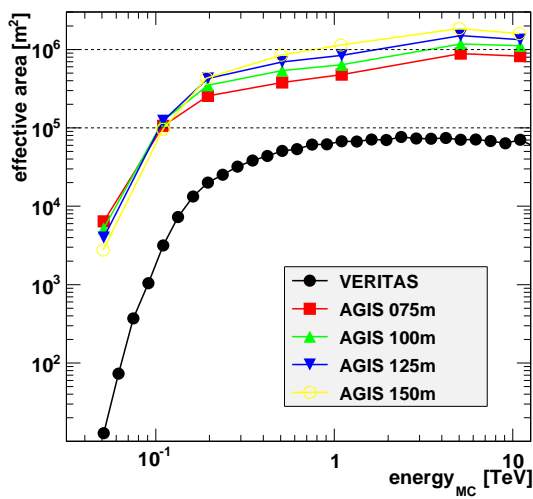}
\caption{Effective area vs energy for a 36 telescope array with different telescope distances
in comparison with VERITAS \cite{Holder-2008}.}
\label{fig:EffArea}
\end{figure}

Figure \ref{fig:EffArea} shows that an effective area of $\approx$1 km$^2$ can be achieved with a 
36-telescope array.
 This is a factor of 10 larger than the effective area of VERITAS and results
in a factor $\approx$3 improvement in sensitivity by itself. The figure also shows that arrays with
small telescope distances ($<$100 m) perform best at energies below 150 GeV, while larger
telescope distances are best in the TeV range. The final array layout of the AGIS
observatory probably will consist of a dense core surrounded by telescopes at larger distances.

\section{Conclusions}

We discussed results from simulations of the next-generation
gamma-ray observatory AGIS. 
It has been
shown that compared to the current instruments, a factor
of ten improvement in effective area and an 
angular resolution of 1-3 arcmin  can be achieved
with the 36 telescope AGIS array.

\end{document}